\documentclass{ws-procs975x65}

\begin{document}

\title{The Diagnostic Power of X-ray emission lines in GRBs}

\author{M. B\"ottcher}

\address{Department of Physics and Astronomy\\
Ohio University\\ 
Athens, OH 45701, USA\\
E-mail: mboett@helios.phy.ohiou.edu}

\maketitle

\abstracts{Absorption and reprocessing of Gamma-ray burst 
radiation in the environment of cosmological GRBs can be used 
as a powerful probe of the elusive nature of their progenitors.
Although it is widely accepted that long-duration GRBs are 
associated with the deaths of massive stars, at least two 
fundamentally different scenarios concerning the final collapse
are currently being considered. Delayed reddened excesses 
in the optical afterglows of several GRBs indicate that 
a supernova, possibly of Type Ic, takes place within a 
few days of a GRB. This supports the collapsar model, where 
the core of a massive star collapses promptly to a black hole.
Variable X-ray features observed in the prompt and afterglow
spectra of several GRBs, suggest that a highly metal enriched 
and dense shell of material surrounds the sources of GRBs. In
some cases, evidence for expansion of these shells with velocities
of a substantial fraction of the speed of light has been claimed. 
These observations have been interpreted as support for the 
supranova model, where a massive star collapses first to a 
supramassive neutron star, which later collapses to a black 
hole following loss of rotational support. In this review paper, 
I will present a brief overview of the current status of the 
observational evidence for X-ray spectral features in GRBs, 
and discuss their implications for both the collapsar and the 
supranova model.
}

\section{\label{intro}Introduction: Summary of observed X-ray 
spectral features in GRBs}

The precise localization of gamma-ray bursts (GRBs) by the BeppoSAX 
satellite, launched in 1996, has facilitated the subsequent discovery 
of X-ray and optical GRB afterglows, the measurement of redshifts of
GRBs and the firm establishment of their cosmological distance scale 
(at least for long GRBs with durations $t_{90} \gtrsim 2$~s) beyond 
any reasonable doubt. The physics of the generally smoothly decaying
radio through X-ray continuum afterglow emission is now believed to be 
rather well understood in terms of the external synchrotron shock model 
(for recent reviews see, e.g., ref.~\refcite{meszaros02} or \refcite{dermer01}.) 
However, in spite of these significant advances, the ultimate source of GRBs is
still a matter of vital debate. This is mainly due to the fact that the 
continuum GRB afterglows are the ``smoking gun'' of the GRB explosion, 
revealing only very little information about the initial energy source. 

However, even without a direct observation of the central engines of 
GRBs, it might be possible to infer their nature indirectly if detailed 
probes of the structure and composition of their immediate vicinity 
can be found. A promising candidate for such a probe is high-quality, 
time resolved X-ray spectroscopy of GRB afterglows. Several high-quality
spectra of GRB X-ray afterglows have been obtained from recent observations
by {\it Chandra} and 
{\it XMM-Newton}\cite{piro00,reeves02,watson02a,watson02b,reeves03,bt03,watson03,butler03}. 
The Swift mission, scheduled for launch early 2004, is expected to give 
another boost to the study of early X-ray afterglows of GRBs. Previous and 
currently operating X-ray telescopes have so far (status: December 2003) 
revealed marginal evidence for X-ray emission lines (mostly consistent 
with Fe~K$\alpha$ fluorescence lines) or radiative recombination edges 
in 7 GRBs, and one case of a transient X-ray absorption feature at an 
energy consistent with an iron K absorption edge. 

In the following, I will give a brief review of the observed X-ray spectral 
features in GRBs, before turning to the implications of those observations 
in general terms. This will be followed by a summary of the currently 
discussed GRB models which could give rise to the conditions required 
to produce the observed spectral features.

\subsection{\label{intro_emission}X-ray emission lines in GRB afterglows}

Marginal evidence for transient X-ray emission line features in
GRB afterglows have been reported for 7 Gamma-ray bursts: 
GRB~970508\cite{piro99}, GRB~970828\cite{yoshida99}, 
GRB~991216\cite{piro00}, GRB~000214\cite{antonelli00}, 
GRB~011211\cite{reeves02}, GRB~020813\cite{butler03}, and
GRB~030227\cite{watson03}. Tab. \ref{line_table} lists the key 
observational features of these seven GRB line detections: (1) the 
GRB designation, (2) the redshift, (3) the time interval of the 
line detection, (4) the isotropic luminosity of the emission line, 
(5) the instrument with which the detection was made, (6) additional
remarks concerning the line detection, and (7) the reference to
the detection and important contributions through re-analysis of 
the respective observations. Note that the beginning of the time
intervals listed in column (3) are generally set by the beginning
of the respective observations, typically several hours -- $\sim 1$~day 
after the GRB, and might thus not be representative of the onset of 
the line emission. Only in the case of GRB~030227\cite{watson03} was
a clean, featureless soft X-ray spectrum obtained by {\it XMM-Newton} 
in one segment of the observation prior to the segments showing evidence
for emission lines. The luminosities quoted in column (4) refer to the
Fe~K$\alpha$ line emission, unless explicitly noted otherwise.

\begin{sidewaystable}	
\tbl{Summary of the observed X-ray emission lines in GRB afterglows.
\label{line_table}}
{\begin{tabular}{@{}ccccccc@{}}
\hline
\multicolumn{7}{c}{}\\[-2ex]
GRB & z & $t_{\rm det}$ [s] & $L_{\rm line}$ [erg/s] & Instrument 
& Remarks & References \\[0.25ex]
\hline
\multicolumn{7}{c}{}\\[-2ex]
970508 & 0.835 & $2 \times 10^4$ -- $5.6 \times 10^4$   & $6 \times 10^{44}$ & BeppoSAX               & second. outburst         & \refcite{piro99} \\ 
970828 & 0.958 & $1.2 \times 10^5$ -- $1.4 \times 10^5$ & $5 \times 10^{44}$ & ASCA                   & RRC without line         & \refcite{yoshida99,yonetoku01} \\ 
991216 & 1.00  & $1.3 \times 10^5$ -- $1.5 \times 10^5$ & $8 \times 10^{44}$ & Chandra ACIS-S + HETG  & Broad line + RRC         & \refcite{piro00} \\ 
000214 & 0.47  & $4 \times 10^4$ -- $1.5 \times 10^5$   & $4 \times 10^{43}$ & BeppoSAX               & No OT; z from X-ray line & \refcite{antonelli00} \\ 
011211 & 2.14  & $4 \times 10^4$ -- $6.7 \times 10^4$   & Si: $6.4 \times 10^{44}$ & XMM-Newton       & No Fe; line det. controversial & \refcite{reeves02,bt03,reeves03,rs03} \\ 
       &       &                                        & S: $6.2 \times 10^{44}$  & \\
020813 & 1.254 & $7.6 \times 10^4$ -- $1.5 \times 10^5$ & Si: $1.1 \times 10^{44}$ & Chandra HETGS    & No Fe; low-$\sigma$ Ni   & \refcite{butler03} \\ 
       &       &                                        & S: $1.6 \times 10^{44}$  & \\
030227 & $\sim 1.6$? & $7 \times 10^4$ -- $\ge 8 \times 10^4$ & Si: $6 \times 10^{44}$ & XMM-Newton   & No host red shift; no Fe or Ni & \refcite{watson03} \\ 
       &       &                                              & S: $4 \times 10^{44}$  & \\
\\[0.5ex]
\hline
\end{tabular}}
\begin{tabnote}
Unless otherwise noted, the luminosity in the 4$^{th}$ column refers to 
the luminosity of the Fe~K$\alpha$ line. The start times indicated in 
the third column generally mark the beginning of the afterglow observation, 
not the actual onset of the line emission. The remarkable exception is 
GRB~030227, where the onset of the line emission was directly observed.
\end{tabnote}
\end{sidewaystable}

In the {\it BeppoSAX} NFI observation of the afterglow of GRB~970508,
Piro et al.\cite{piro99} detected evidence, at the $\sim 3 \, \sigma$ level,
for an emission line feature consistent with a 6.7~keV Fe~K$\alpha$ 
line from highly ionized iron, at the redshift of the burst at 
$z = 0.835$. The X-ray afterglow of this burst exhibited a secondary
outburst after $\sim 6 \times 10^4$~s, and the disappearance of the
line from the X-ray spectrum seemed to be coincident with the onset of
this secondary X-ray outburst. From a re-analysis of the same observational
data, Paehrels et al.\cite{paehrels00} concluded that a fit using a
plasma emission model in photoionization equilibrium, assuming
an illuminating continuum identical to the GRB afterglow emission, 
would either require too large a redshift or too high a temperature
for a photoionization-dominated model. This was interpreted as possible 
evidence for thermal plasma emission rather than emission from a 
photoionized plasma.

Interestingly, the second GRB showing evidence for an X-ray emission
feature at the redshifted Fe~K$\alpha$ energy, GRB~990828, appears to 
be opposite in that respect. Before the redshift of GRB~970828 could 
be determined, Yoshida et al.\cite{yoshida99} inferred a redshift of 
$z = 0.33$ from the energy of the line feature in the {\it ASCA} spectrum 
of the afterglow of this GRB, assuming that it corresponds to an Fe~K$\alpha$ 
line at 6.4~keV in the burst rest frame. In a later re-analysis, after 
the likely host-galaxy identification of GRB~970828, associated with a 
redshift of $z = 0.958$\cite{djorgowski01}, Yoshida et al.\cite{yoshida01} 
and Yonetoku et al.\cite{yonetoku01} argued that the emission feature is 
consistent with this redshift if it is a radiative recombination continuum 
(RRC) edge rather than a fluorescence or recombination line. They argue 
that the electron temperature in the plasma responsible for the emission 
feature is inconsistent with the ionization temperature in thermal equilibrium, 
and would therefore indicate photoionization as the dominant ionization
mechanism.

Marginal evidence for a RRC was also reported for the {\it Chandra} 
ACIS-S+HETG spectrum of the afterglow of GRB~991216\cite{piro00}. 
Piro et al. (2000) interpret the width and the apparently blue-shifted 
best-fit energy of the iron line with respect to the redshift of the 
host galaxy at $z = 1.00$ as evidence that the Fe line + RRC originate 
in an outflow with a velocity of $v \sim 0.1 \, c$. If this interpretation 
is correct, it might be indicative of a supernova explosion a few months 
prior to the GRB. 

For GRB~000214, no optical counterpart could be identified. In this
case, the only information concerning its redshift comes from the
X-ray emission line detected by the {\it BeppoSAX} NFI, and was
estimated to be $z = 0.47$, if the emission line at $E = (4.7 \pm
0.2)$~keV is interpreted as the redshifted Fe~K$\alpha$ line from
hydrogen-like iron\cite{antonelli00}.

While up to GRB~000214, only evidence for emission features
attributable to Fe (or possibly Ni and/or Co) had been detected
in GRB afterglows, the observational picture seems to have changed
dramatically with more regular GRB afterglow observations by {\it Chandra}
and especially {\it XMM-Newton}. Reeves et al.\cite{reeves02,reeves03} 
reported the marginal detection of a set of emission lines in the 
{\it XMM-Newton} spectrum of the early afterglow of GRB~011211. If 
real, these features would be peculiar in that they show evidence 
for K$\alpha$ lines from the hydrogen-like ions of lighter elements, 
such as Si~XIV, S~XVI, Ar~XVIII, and CA~XX, but no indication of an 
Fe~K$\alpha$ line or RRC edge. Those lines appear to be blue-shifted 
by an average of $v \sim 0.1 \, c$ with respect to the likely redshift 
of the burst, $z = 2.14$. However, Borozdin \& Trudolyubov\cite{bt03}
have pointed out several caveats of the tentative line detections
in GRB~011211. (1) The set of low-Z metal emission lines appeared
only in the first $\sim 5$~ksec, prior to a re-pointing of the
spacecraft. During that time, the source was located very close 
to the edge of the CCD chip of the EPIC-pn detector. (2) Evidence 
for the lines appears only in the EPIC-pn detector, with no such 
indication in the EPIC-MOS detectors. (3) In their analysis, 
Borozdin \& Trudolyubov\cite{bt03} find the {\it XMM-Newton} 
spectrum of GRB~011211 consistent with a pure power law, with 
no evidence for spectral evolution in consecutive time intervals. 
The addition of an emission line system lowers the reduced 
$\chi^2$ from 1.03 to 0.82 (with 8 additional free parameters), which 
might be an indication of over-fitting, rather than a statistically 
significant improvement of the fit. A statistical analysis based on 
a Monte-Carlo simulation approach by Rutledge \& Sako \cite{rs03} 
yielded a blind-search confidence level of only $\sim 77$ -- 82~\% 
if the line energies are not restricted to a narrow range around 
the host-galaxy redshift of $z = 2.14$. However, this may have 
been the result of a coarser energy binning than used by Reeves 
et al.\cite{reeves02,reeves03}, through which much of the line 
structure may have been lost. Reeves et al.\cite{reeves03} provided
a detailed response to the concerns mentioned above, finding that 
the results of their initial analysis are robust, but the reliability 
of the line detection remains controversial to date.

A second example of a set of low-Z element K$\alpha$ emission lines
was found in the early afterglow of GRB~020813\cite{butler03}. The
most prominent lines were identified with K$\alpha$ fluorescence line
emission from Si~XIV and S~XVI, blueshifted by $v = 0.12$~c with
respect to the host galaxy red shift of $z = 1.255$\cite{barth03}. 
The significance of these line detections is quoted at $\gtrsim 2.9 \, 
\sigma$. No Fe line was detected, but weak evidence for Ni emission 
was found.  

\begin{figure}[ht]
\centerline{\epsfxsize=3.8in\epsfbox{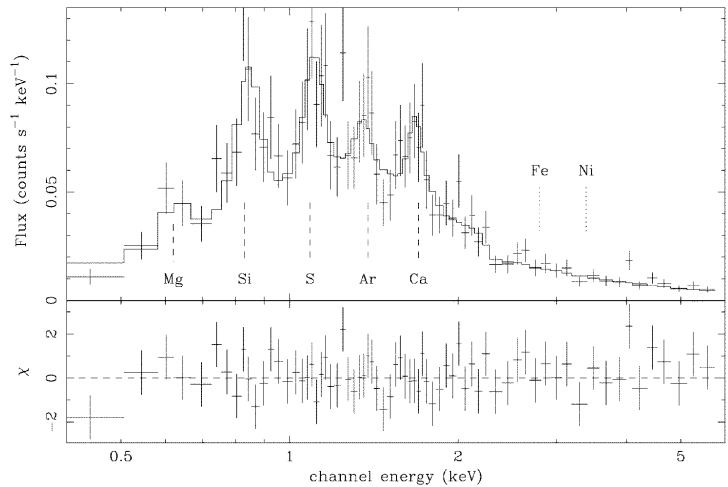}}
\caption[]{\label{fig030227}EPIC-pn spectrum of the last 
11 ks of the {\it XMM-Newton} observation of the afterglow 
of GRB~030227; from Watson et al.\cite{watson03}}
\end{figure}

The most recent case (at the time of writing: mid-Dec. 2003) of an emission 
line detection in a GRB afterglow, namely in GRB~030227\cite{watson03}, is 
remarkable in at least two aspects. First, it provided the highest level of
confidence of an X-ray line detection to date. Second, it was the first time
that the actual onset of the line emission could be determined. As in the
case of GRB~011211 and GRB~020813, emission lines could be attributed only 
to low-Z elements, specifically, Mg, Si, S, Ar, and Ca (see Fig.~\ref{fig030227}). 
The most significant detections were reported from Si~XIV and S~XVI at red 
shifts of $z = 1.32$, and $z = 1.34$, respectively. To date, no independent 
measurement of the host galaxy red shift has been reported. If the lines are 
identified with Si and S recombination line emission from material flowing 
out at $v \sim 0.1$~c, as indicated in other cases, then a source red shift 
of $z \sim 1.6$ could be inferred. No lines from iron-group elements were 
detected. GRB~030227 was the first case of a GRB detected by the burst 
alert system on board {\it INTEGRAL} to be localized to within a few arc
seconds. Note that Mereghetti et al.\cite{mereghetti03} had previously 
analyzed the time-averaged {\it XMM-Newton} spectrum of this observation,
in combination with simultaneous {\it INTEGRAL} observations, and found
marginal ($\sim 3.2 \, \sigma$) evidence for an X-ray line at 1.67~keV,
which would be coincident with the Ca~XX K$\alpha$ line identified by
Watson et al.\cite{watson03}. 

\begin{figure}[ht]
\centerline{\epsfxsize=3.8in\epsfbox{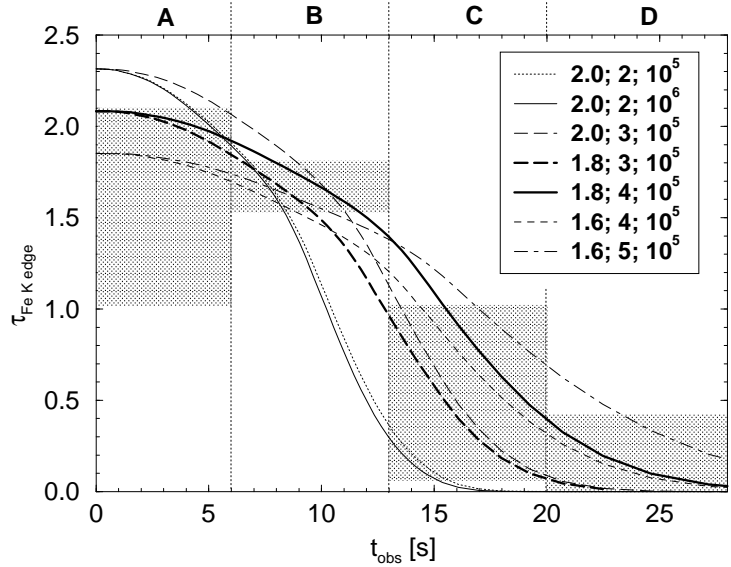}}
\caption[]{\label{fig990705}Time evolution of the depth of the
transient absorption edge in GRB~990705\cite{amati00}. The curves
indicate fits with a time-dependent photoelectric absorption and
photoionization model in a homogeneous shell. The curves are labelled
by (column density of material in the shell, $N_H$ [$10^{22}$~cm$^{-2}$],
assuming 75-fold overabundance of iron; inner radius of the absorbing 
shell $r_{\rm in}$ [$10^{18}$~cm]; total density of material in the 
shell, $n$ [cm$^{-3}$]).}
\end{figure}

\subsection{\label{intro_absorption}The Transient Absorption Feature 
in GRB~990705}

Atomic X-ray absorption features in GRB afterglows are rather common.
However, in order to distinguish such features from foreground
absorption and to associate them physically with the GRB, one 
needs to find evidence for variability of those absorption features 
on the GRB / early afterglow time scale\cite{pl98,boettcher99}.
Until now, only one such case has been observed: In the prompt
X-ray emission of GRB~990705, a transient absorption feature 
has been found at $(3.8 \pm 0.3)$~keV\cite{amati00}, consistent 
with an Fe~K absorption edge from neutral iron at $z = 0.86 \pm 
0.17$ at the redshift of the host galaxy at $z = 0.84$\cite{lazzati01}.
This absorption edge was only seen in the first 13~s of the GRB, 
while no evidence for excess absorption was found in later 
segments of the {\it BeppoSAX} WFC observations (see Fig.~\ref{fig990705}). 
The best fit to the segment with the most significant detection of 
the absorption edge (6 -- 13~sec.), assuming an underlying power-law
absorbed by a neutral absorbing column $N_H$ with solar abundances
and an additional absorption edge yielded a depth of the absorption 
feature of $\tau = 1.4 \pm 0.4$ and $N_H = (3.5 \pm 1.4) \times
10^{22}$~cm$^{-2}$. Leaving the iron abundance in the neutral absorber 
as a free parameter in a power-law fit to the spectrum in this time 
segment resulted in $N_H = (1.32 \pm 0.3) \times 10^{22}$~cm$^{-2}$
and a relative overabundance of iron with respect to solar abundances
of $X_{\rm Fe} = 75 \pm 19$. Implications and possible interpretations
of these results will be discussed in the final section of this review.

\section{\label{general_constraints}General Constraints}

Analytic estimates of the fluorescence and/or recombination line 
emission in photoionized media in GRB environments have been 
presented even before their actual detection by several authors
(e.g., Refs. \refcite{mr98,ghisellini99,lazzati99,boettcher00}).
A very general estimate of the amount of iron required to produce 
the observed iron line features can be derived from considering 
that in the course of complete photoionization of iron, on average 
$\sim 5$ Fe~K$\alpha$ photons are emitted, since it takes on average 
$\sim 12$ X-ray photons to ionize an initially neutral iron atom 
completely (taking into account the Auger effect), and the K$\alpha$ 
fluorescence yield is in the range 0.3 -- 0.4 for the various ionization 
stages of iron. In dilute media, if recombination is negligible, it 
would then take $N_{\rm Fe} \sim 2 \times 10^{56} L_{44} \, t_5$ iron 
atoms to produce an iron line of isotropic luminosity $L_{\rm line} = 
10^{44} \, L_{44}$~ergs~s$^{-1}$ over a time scale of $\Delta t = 10^5 
\, t_5$~s. If recombination and multiple ionization of the same iron 
atom enhance the efficiency of line production, one can introduce an 
ehnancement factor $f$, counting the number of times that a single iron 
atom can effectively contribute 5 K$\alpha$ photons in the process of 
repeated cycles of ionization and recombination. One then arrives at 
a general estimate of

\begin{equation}
M_{\rm Fe} \approx 0.16 \, {L_{44} \, t_5 \over f} \; M_{\odot}
\label{M_Fe_general} 
\end{equation}
In dilute, extended media, where recombination is inefficient, $f \sim 1$,
while in dense media with efficient recombination, $f \gg 1$. Due to light
travel time effects, the duration of the line emission will be at least
$t_{\rm line} > R \, (1 + z) \, (1 - \cos\theta_{\rm obs}) /c$, where $R$ 
is the extent of the reprocessing material. Thus, the size limit, together 
with the restriction that in the GRBs with line detections, there is no 
indication for excess X-ray absorption, leads to typical mass estimates 
of $M_{\rm Fe} \sim 0.1$ -- 1~$M_{\odot}$ of iron, confined in $R \lesssim 
10^{-3}$~pc, if the line emission were to originate in a dilute, 
quasi-isotropic environment. This is unlikely to be realized in 
any astrophysical setting, and may thus be ruled 
out\cite{ghisellini99,boettcher99}. Consequently, inhomogeneous 
media with significant density enhancement outside our line of 
sight toward the GRB source are required\cite{lazzati99,boettcher00}.

\begin{figure}
\begin{center}
\includegraphics[width=14cm]{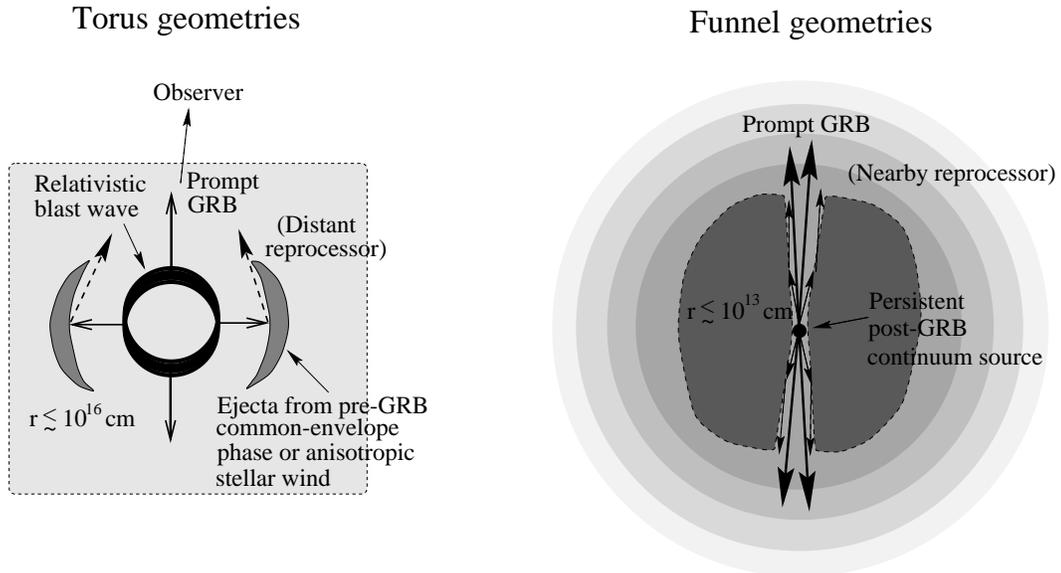}
\caption{\label{reprocessors}Fundamental geometries of distant 
reprocessors (left) and nearby reprocessors (right).}
\end{center}
\end{figure}

At this point, several geometries and general scenarios will have to be 
distinguished. First, the duration $t_{\rm line}$ of the line detection
can be set either by the light travel time effect, which puts the
reprocessor at distances of $10^{15} \, {\rm cm} \lesssim R \lesssim
10^{16}$~cm. These types of configurations are referred to as {\it
distant reprocessor models}. Alternatively, the duration of the line
emission can be determined by the duration of the illumination by a
more persistent, gradually decaying central source. In that case, the
reprocessor can be located much closer to the source than in the case
of the distant reprocessor models. For this reason, these scenarios
are referred to as {\it nearby reprocessor models} (see Fig.~\ref{reprocessors}). 
Second, we need to distinguish between different mechanisms ionizing 
the line-emitting iron atoms. The scenarios considered above are 
generally based on photoionization being the dominant ionization 
mechanism. However, it is very well conceivable that the material 
in the vicinity of GRBs is energized by shocks associated with the 
GRB explosion, and heated to temperatures of $T \sim 10^7$ -- $10^8$~K 
(e.g., Ref.~\refcite{vietri99}). In this case, and for sufficiently high 
densities, collisional ionization may dominate over photoionization, 
and the iron K$\alpha$ line emission will be dominated by the 
recombination lines of H and He-like iron, accompanied by a radiative 
recombination continuum. We will refer to these models as ``thermal 
models'', as opposed to the ``photoionization models'' mentioned above. 

\section{\label{photoionization}Photoionization Models}

General parameter studies on photoionization models have been done
by many authors. An important complication to keep in mind is the
fact that the continuum illuminating the reprocessing material
might not be identical or even similar to the observed afterglow 
emission because, as explained above, the reprocessor has to be
located off the line of sight towards the central source of the
GRB. As it seems well-established today that GRB emission (at
least from long-duration GRBs) is beamed, any emission off the
main cone in which a GRB is observed might be different from the
actual GRB and afterglow radiation. Furthermore, if the reprocessor 
is located at a distance of $R = 10^{15} \, R_{15}$~cm, and 
illuminated by GRB (and early afterglow) emission from a 
relativistic blast wave advancing at a coasting speed 
corresponding to $\Gamma = 100 \, \Gamma_2$, then it will 
be swept up by the blast wave after a typical illumination 
time of

\begin{equation}
t_{\rm ill} \sim 2 \, {R_{15} \over \Gamma_2^2} \; {\rm s}
\label{t_ill}
\end{equation}
unless the blast wave deceleration radius $r_{\rm dec} = 5 \times
10^{16} \, (E_{52} / n_0 \, \Gamma_2^2)^{1/3}$~cm is much less than
$10^{15}$~cm. Here, $E_{52}$ is the isotropic equivalent energy of
the GRB explosion in units of $10^{52}$~ergs~s$^{-1}$ and $n_0$ is 
the density of the (homogeneous) surrounding medium in units of 
cm$^{-3}$. The condition $r_{\rm dec} \ll 10^{15}$~cm would require
a very low explosion energy directed toward the reprocessor, a large
density of the decelerating external medium, and/or a very high bulk
Lorentz factor $\Gamma$, i. e. a very low baryon contamination.

General studies of the reprocessing efficiency and spectral and
temporal signatures of reprocessor models for X-ray emission lines
in GRB environments\cite{brr01,lazzati02,ghisellini02,kallman03}
have yielded very useful insight into the general properties of such 
reprocessing features. The above caveat needs to be taken into 
account very carefully when applying these results to real GRBs 
and scaling illuminating spectra, ionization parameters etc., to 
properties derived from observed continuum afterglows of GRBs. 

In a parameter-independent study in Ref.~\refcite{lazzati02}, the 
efficiency of reprocessing a given ionizing flux into fluorescence
and/or recombination line flux has been calculated as a function 
of ionization parameter. This efficiency is generally at most 
$\sim 1$~\% and dependent on the precise spectral shape of the 
illuminating spectrum. This implies that the observed iron lines 
listed in Tab. \ref{line_table} require a total energy in the 
ionizing continuum of at least $\sim 10^{51}$~ergs\cite{ghisellini02}. 
The reprocessing efficiency is drastically decreasing for high 
ionization parameters $\xi \gtrsim 10^3$. This may provide a 
diagnostic between distant and nearby reprocessor models\cite{lazzati02}.

Another important aspect associated with iron K$\alpha$ line emission
is related to radioactive decay of ${^{56}{\rm Ni}} \to {^{56}{\rm Co}} 
\to {^{56}{\rm Fe}}$ in the ejecta of a supernova which may be associated 
with the GRB\cite{lazzati01,mclaughlin02,mlw02}. Since it is primarily 
$^{56}$Ni that is produced in the final stages of a supernova progenitor, 
any emission line features should gradually shift in energy from 8~keV 
for Nickel to 7.4~keV for Cobalt and 6.9~keV for hydrogen-like iron on 
the radioactive decay time scales of 6.1~d and 78.8~d for $^{56}$Ni 
and $^{56}$Co, respectively. However, McLaughlin et al.\cite{mclaughlin02} 
and McLaughlin \& Wijers\cite{mlw02} point out the curious detail 
that $^{56}$Ni would usually predominantly decay via electron 
capture. Thus, if the reprocessing material happens to be
highly ionized, the decay time scale may be significantly longer
than in the standard case of neutral $^{56}$Ni. Consequently, for
detailed modeling of the ``iron'' line emission in GRB afterglows, 
a rest-frame line energy of $\sim 8$~keV might actually be a more 
appropriate assumption than the standard value of 6.4 -- 6.97~keV. 
In the remainder of this review, I will, for simplicity, continue 
to refer to any line feature around $E \sim 7$~keV in the GRB rest 
frame as ``iron line'', keeping in mind that it might have a 
significant contribution from Ni and Co.

In addition to light-travel time effects, standard constraints on 
reprocessor models are obtained considering the ionization and
recombination time scales. The ionization time scale may be estimated
as

\begin{equation}
t_{\rm ion} \approx {12 \over \int_{E_{\rm thr}}^{\infty} 
\Phi_{\rm ion} (E) \, \sigma_{\rm pi} (E) \, dE} \approx 
{12 \, (\alpha + 2) \over \sigma_0 \, \Phi_0 \, E_{\rm thr}},
\label{t_ion}
\end{equation}
where $E_{\rm thr}$ is the photoionization threshold energy,
$\sigma_{\rm pi} (E) \approx \sigma_0 \, (E / E_{\rm thr})^{-3}$
is the photoionization cross section, and $\Phi(E) = \Phi_0 \,
(E / E_{\rm thr})^{-\alpha}$ is the ionizing photon flux, assumed to 
be a straight power-law above the ionization threshold. Appropriate
averages for the various ionization stages of iron are $E_{\rm thr} 
\sim 8$~keV and $\sigma_0 \sim 3.5 \times 10^{20}$~cm$^{-2}$. The 
recombination time scale can be approximated as
\begin{equation}
t_{\rm rec} = {1 \over n_e \, \alpha_{\rm rec}} \approx {10^9 \, T_4^{3/4} 
\over n_H \, (Z_{\rm eff} / 26)^2} \; {\rm s},
\label{t_rec}
\end{equation}
where $\alpha_{\rm rec}$ is the recombination coefficient, $Z_{\rm eff}$
is the effective nuclear charge, $T_4$ is the electron temperature in 
units of $10^4$~K, and $n_H$ is the hydrogen number density in units
of cm$^{-3}$. The approximation for the recombination rate used above
assumes that recombination is dominated by radiative recombination,
which becomes inaccurate for ionization states lower than Fe~XXV at
temperatures above $\sim 10^5$~K. 

Additional model constraints are derived from considerations concerning 
optical-depth effects due to resonance scattering out of the line of 
sight (for the resonant Ly$\alpha$ line of Fe XXVI) and Thomson 
scattering, which define a maximum depth in the reprocessor beyond
which the material will effectively no longer contribute to the line
emission. Parameter constraints for reprocessor models have recently
been summarized in Ref.~\refcite{tgl03} for the specific purpose of 
modelling the low-Z element line detections in GRB~011211.

\subsection{\label{distant}Distant Reprocessor Models}

Distant reprocessor models, in which the duration of the line
emission is dominated by the light travel time difference, 
$t_{\rm line} \sim R \, (1 + z) \, (1 - \cos\theta_{\rm obs}) / c$, were
the first to be discussed after it became clear that quasi-isotropic
fluorescence line emission scenarios were infeasible to explain the
observed iron line features\cite{lazzati99,boettcher00}. 
Generally, in these models, a photoionizing continuum from 
the central source, emitted in tandem with the prompt and early
afterglow radiation, is impinging on the surface of a strongly
anisotropic configuration of dense, pre-ejected material. 
Recall, however, that the ionizing continuum does by no means have 
to be identical or even similar to the observed GRB and afterglow 
emission. The geometry of the reflector is generally taken to be
a torus, or a quasi-uniform shell with an evacuated funnel through 
which the prompt GRB emission can escape unscattered\cite{weth00}. 
The pre-ejected tori could plausibly be the result of a 
common-envelope phase preceding the GRB event. All types of
GRB models pertaining to the class of black-hole accretion-disk
models, including the collapsar/hypernova and the He-merger (see, 
e.g., ref.~\refcite{fwh99} for a review) are likely to have
undergone a common-envelope phase prior to the primary's core
collapse. The material ejected during such a common-envelope
phase is expected to have a directed velocity of the order of
the escape velocity from the secondary, $v \sim \sqrt{2 G M_{\rm sec} 
/ R_{\rm sec}} \sim 6 \times 10^7 \, (m/r)^{1/2}$~cm~s$^{-1}$, where
$M_{\rm sec} = m \, M_{\odot}$ and $R_{\rm sec} = r \, R_{\odot}$
are the secondary's mass and radius. For a detailed discussion of the 
expected structure of pre-ejected disks / tori, see ref.~\refcite{bf01}
and references therein. In the collapsar/hypernova 
models, the delay between the ejection of the primary's hydrogen 
envelope and the GRB may be as large as 100,000~years, although 
significant uncertainties about the actual delay time scale remain. 
Such a long delay would place the pre-ejected material at radii
$R \gg 10^{16}$~cm, probably too large to be consistent with 
the observed time delays of the GRB X-ray emission line features. 

Much smaller delays are expected in the He-merger 
scenario\cite{zf01} and the supranova model\cite{vs98}. 
In the He-merger model, time delays of a few hundred years to a 
few times 10,000 years may be typical, allowing for both nearby
and distant reprocessor scenarios. The supranova model predicts
delays of the order of the spin-down time scale of the supramassive
neutron star, $t_{\rm sd} \sim 10 \, j_{0.6} \, \omega_4^{-4} \, 
B_{12}^{-2}$~yr, where $j_{0.6}$ is the angular-momentum parameter
in units of 0.6, $\omega_4$ is the angular velocity in units of 
$10^4$~s$^{-1}$, and $B_{12}$ is the surface magnetic field in 
units of $10^{12}$~G. Since the typical magnetic field strength 
is very poorly constrained, delays of several months to several 
thousands of years might be possible for this model. The ejecta
velocity in this case should be more typical of supernova ejecta, 
$v \sim 10^9$~cm~s$^{-1}$. Thus it appears that the supranova model 
may be able to accomodate both distant and nearby reprocessor models 
as well as the thermal models discussed in the following section. 

Different geometrical variations of distant reprocessor models 
have been discussed in more detail in Refs.~\refcite{lazzati99,vietri01}.
Analytical estimates as well as detailed numerical simulations 
of distant reprocessor scenarios\cite{boettcher00,weth00} indicate 
a mass requirement of $M_{\rm fe} \sim 10^{-5}$ -- $10^{-4} \, 
M_{\odot}$ for most of the observed Fe~K$\alpha$ emission 
line features observed to date. Note, however, that Vietri 
et al.\cite{vietri01} derive a significantly higher mass 
estimate of $M_{\rm Fe} \sim 1 \, M_{\odot}$ for GRB~991216, 
the most extreme case in terms of total energy emitted in 
the Fe~K$\alpha$ line.

\subsection{\label{nearby}Nearby Reprocessor Models}

Nearby reprocessors, at typical distances of $\lesssim 10^{13}$~cm,
can either consist of pre-GRB ejecta (see discussion above) or the
expanding envelope of a super-giant GRB progenitor\cite{rm00,mclaughlin02}.
In the latter case, if the energy release in the GRB is strongly 
beamed, the envelope of the progenitor star is expected to remain 
in a quasi-stable state for a time of the order of the sound-crossing 
time through the progenitor, $t_{\rm sc} \sim 30 \, m_1^{-1/2} \, 
R_{13}^{3/2}$~d, where $m_1$ is the progenitor mass in units of 
$10 \, M_{\odot}$, and $R_{13}$ is its radius in units of 
$10^{13}$~cm. It can thus plausibly survive the break-through 
of an ultrarelativistic jet, and provide a scattering funnel 
for persistent ionizing radiation throughout the prompt and 
early afterglow phase of the GRB. 

In this class of models, the highly collimated ultrarelativistic 
outflow most probably associated with the prompt GRB and the
continuum afterglow emission, has moved far past the nearby
reprocessor at the time the observed emission lines are
produced, and can obviously not contribute to the illuminating
continuum. Thus, these models require a persistent source of 
ionizing radiation over at least the duration over which the
iron line is observed. Rees \& M\'esz\'aros\cite{rm00} suggest 
that such a source could be provided by the gradually decaying 
energy flux from a magnetically driven relativistic wind from
a fast-rotating, strongly magnetized neutron star (magnetar), 
if the primary GRB mechanism does not result in the formation
of a black hole. They arrive at a required mass of $M_{\rm Fe}
\sim 10^{-8} \, M_{\odot}$ of iron in a very thin, ionized
skin of the funnel in order to reproduce an iron line with
properties as observed in GRB~991216. However, if a similar
abundance of iron is distributed throughout the entire stellar
envelope, then a mass estimate more in line with the distant
reprocessor models ($M_{\rm Fe} \sim 10^{-5}$ -- $10^{-4} \,
M_{\odot}$) results. These estimates could be reduced if the 
line emission originates in a dense medium with relativistic 
electron densities of $n_{e, rel} \sim 10^{10}$ -- 
$10^{11}$~cm$^{-3}$. In such an environment, the line emission 
could be enhanced by the \v Cerenkov effect, as pointed out by 
Wang et al.\cite{wang02}. 

\begin{figure}
\begin{center}
\includegraphics[width=8cm]{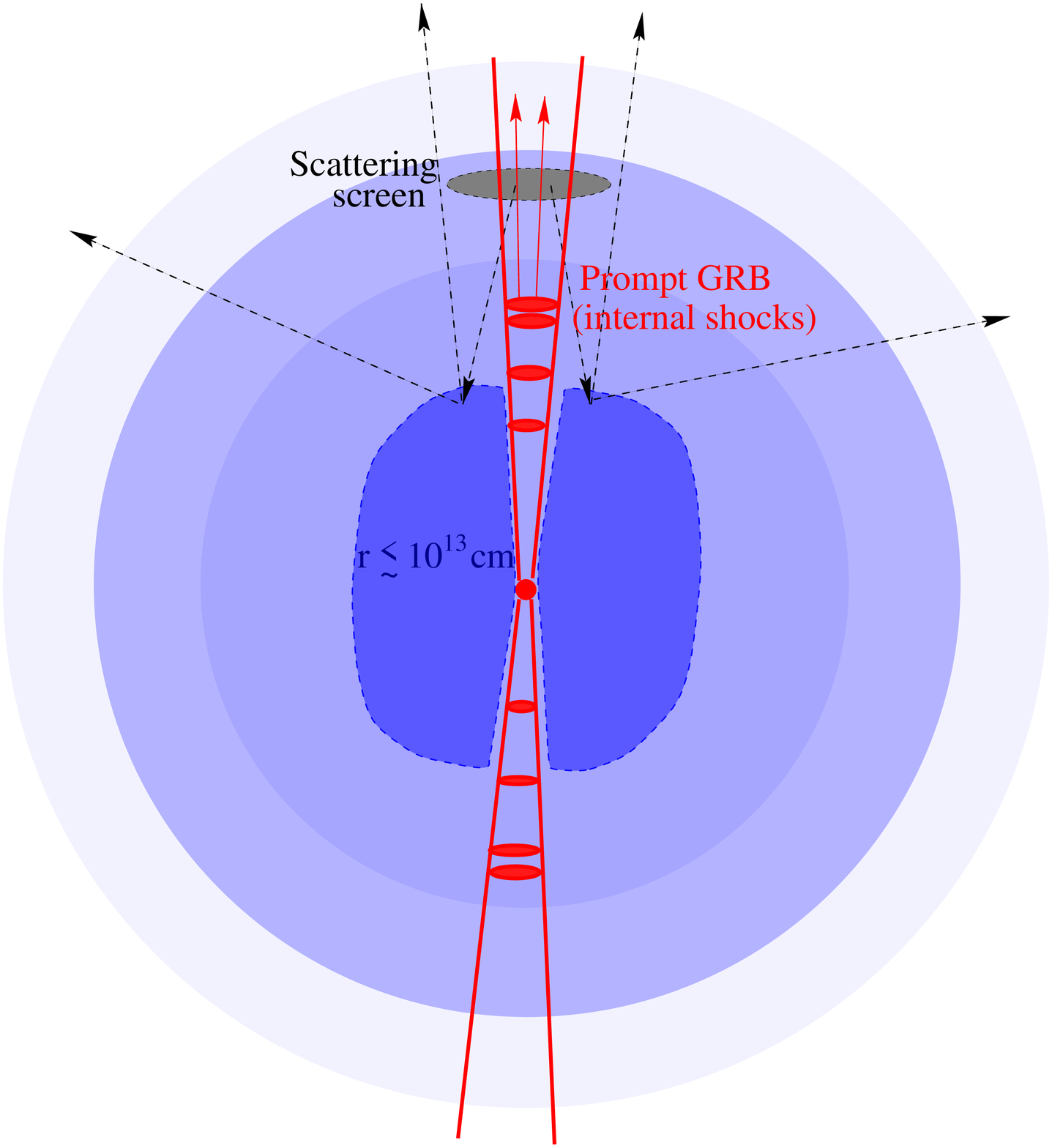}
\end{center}
\caption[]{Geometry of the double-scattering 
model of Refs.~\refcite{vietri01,kn03}.\label{screen}}
\end{figure}

Another variation of nearby reprocessor models has recently been 
proposed by Kumar \& Narayan\cite{kn03}. They suggest the formation 
of a scattering screen at a distance of $\sim 10^{14}$ -- $10^{15}$~cm 
from the GRB source, possibly as a result of $\gamma\gamma$ pair 
creation on back-scattered GRB radiation\cite{tm00,mt00}. GRB 
emission from later (internal) shocks traveling along the jetted 
ejecta, are scattered back onto the outside of the stellar envelope 
of the progenitor (the supernova ejecta), and thus provide the 
ionizing flux for fluorescence line emission (see Fig.~\ref{screen}). 
This model requires a smaller iron content in the ejecta than 
previous models since it utilizes a larger fraction of the surface 
area of the progenitor's stellar envelope. A possible independent 
diagnostic of such a model might be delayed high-energy emission 
due to hadronic interactions when the GRB blast wave plows into 
the scattering screen (see, e.g., ref.~\refcite{katz94}). Kumar 
\& Narayan\cite{kn03} discuss this model particularly in light 
of the low-Z emission lines in GRB~011211\cite{reeves02}. In 
this specific case, their model would require a relatively large 
pre-GRB outflow rate, producing a density in the surrounding medium 
of $n_0 \sim 7 \times 10^7$~cm$^{-3}$, leading to an external-shock 
deceleration radius of $r_{\rm dec} \lesssim 5 \times 10^{14}$~cm. 
This would correspond to an observed deceleration time scale of 
$t_{\rm dec} \lesssim 0.8 \, \Gamma_2^{-2}$~s, which seems to be 
in conflict with the duration of $t_{\rm dur} \sim 270$~s of GRB~011211, 
with no indication of rapid variability on the time scale of the order 
of $t_{\rm dec}$. 

\section{\label{thermal}Thermal Models}

As an alternative to photoionization models, Vietri et al.\cite{vietri99}
had suggested a thermal model in the framework of the supranova
model. They argue that a relativistic fireball associated with the
GRB might hit the pre-GRB supernova remnant within $\sim 10^3$~s
and heat the ejecta to $T \sim 3 \times 10^7$~K. At such temperatures,
the plasma emission is expected to show strong thermal bremsstrahlung
emission as well as line emission, in particular strong Fe~K$\alpha$
recombination line emission. They suggest that the bremsstrahlung and 
recombination continuum may explain the secondary X-ray outburst
observed in GRB~970508. Since the supranova model seems to be consistent
with a large range of SN -- GRB delays, one might expect that secondary
X-ray outbursts and delayed X-ray emission line features on a variety
of time scales can be explained with this type of models. General
constraints on thermal emission scenarios for Fe~K$\alpha$ lines
have also been considered in Ref.~\refcite{lazzati99}.

Generic thermal emission models have been applied successfully to
the recent observations of lower-Z element emission lines. In
all cases, the fits using a collisionally ionized plasma model
require a significant over-abundance of low-Z metals compared to
solar abundances, with upper limits on the Fe, Ni, and Co 
abundances lower than the low-Z abundance best-fit 
values\cite{reeves02,reeves03,butler03,watson03}. In the case
of GRB~030227\cite{watson03} (the most significant low-Z element 
line detection so far), light metal abundances of $\ge 24$ times
solar have been found, while upper limits on the Fe and Ni abundances
of 1.6 and 18 times solar abundance were found. Possible implications
of these results will be discussed in \S~\ref{discussion}.

B\"ottcher \& Fryer\cite{bf01} have investigated the thermal 
X-ray emission from shock-heated pre-ejected material in alternative 
progenitor models (i.e. other than the supranova model), such as 
the collapsar/hypernova and the He-merger model. They found that 
the He-merger scenario provides a feasible setting for the production 
of transient Fe~K$\alpha$ line emission in the range of luminosities 
and durations observed. Since both the He-merger and the 
hypernova/collapsar models are also consistent with much 
larger SN -- GRB delays, they predict that many GRBs may 
display late-time, thermal X-ray flashes from the shock-heating
of pre-ejected material from a common-envelope phase, which could be
detected out to redshifts of $\sim 1$ with currently operating X-ray
telescopes.

An alternative scenario based on thermal emission has been suggested 
by M\'esz\'aros \& Rees\cite{mr01}. As the collimated outflow from the 
central engine of a collapsar is piercing through the stellar
envelope of the progenitor, a substantial amount of energy is deposited
into the stellar material, which might be highly magnetized. After the
jet breaks out of the stellar envelope, this plasma bubble becomes buoyant
and emerges through the evacuated funnel of the envelope within $\sim
10^4$ -- $10^5$~s. At that time, it may have attained a temperature of
$\sim 10^6$ -- $10^7$~K, sufficient to produce iron lines, but potentially
also a plasma emission spectrum dominated by lower-energy lines, as
possibly observed in GRB~011211. 

\begin{figure}
\begin{center}
\includegraphics[width=13cm]{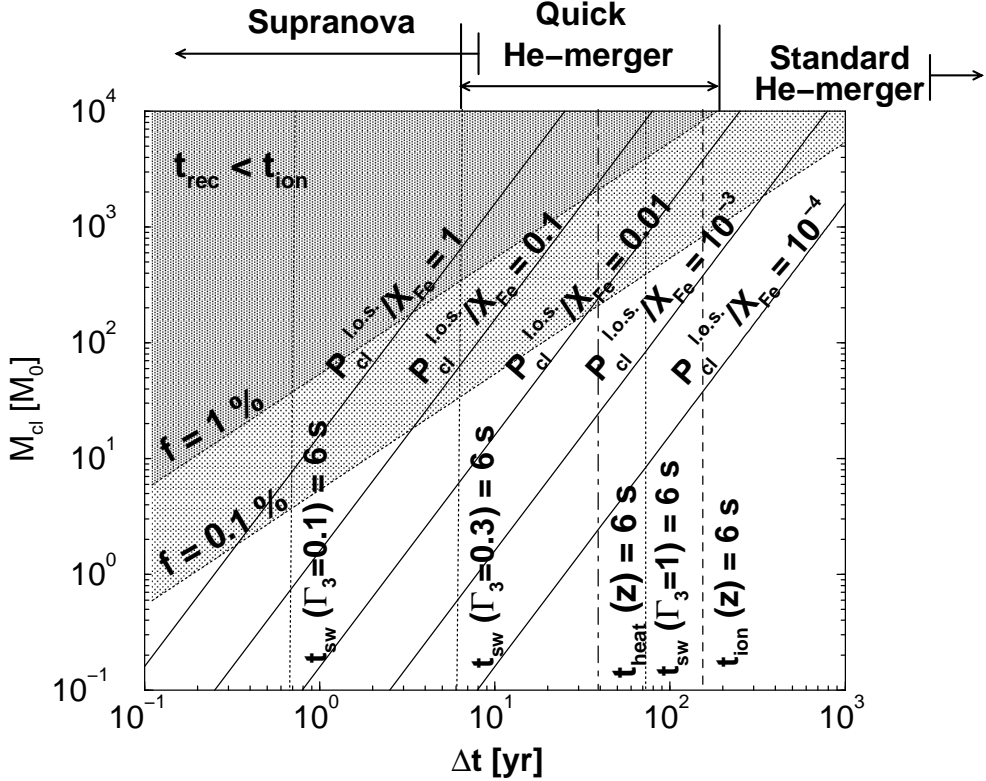}
\caption{Parameter constraints concerning supernova ejecta mass
$M_{\rm cl}$ concentrated in dense clumps, and time delay $\Delta t$
between the primary's supernova explosion and the GRB. Solid lines 
indicate the condition that an Fe~K absorption edge of the depth
observed in GRB~990705 is produced for various values of the ratio
of the probability $P_{\rm cl}^{\rm l.o.s.}$ of an absorbing cloud
being located in the line of sight, to the iron enhancement, $X_{\rm Fe}$,
with respect to standard solar system values. Constellations which would
give a consistent physical scenario must either be located close to
the vertical line corresponding to $t_{\rm ion} (z) = 6$~s (if
recombination is inefficient) or within the shaded regions in the
upper left corner of the plot, which indicates the condition $t_{\rm rec}
\le t_{\rm ion}$ for volume-filling factors of the SN ejecta of
1~\% and 0.1~\%, respectively, 1~year after the SN.}
\end{center}
\label{abs_parameters}
\end{figure}

\section{\label{implications_990705}Implications of the absorption 
feature in GRB~990705}

In principle, inferring constraints on parameters of the circumburster
material from observed GRB properties is an easier task than inferring
them from emission lines, because in the case of absorption features
the continuum responsible for photoionization is identical to the
observed GRB and afterglow continuum. Time-dependent X-ray absorption 
features had been studied for generic, quasi-homogeneous environments 
in Refs. \refcite{boettcher99,ghisellini99}, and for more general 
cases, including radial gradients, in Refs. \refcite{lpg01,lp02}.
For given values of the depth $\tau_{\rm edge}$ of an absorption 
edge and the time scale $t_{\rm edge}$ within which it is 
disappearing, one can directly infer a characteristic radius 
(by setting $t_{\rm ion} = t_{\rm edge}$). Combined with the column
density derived from $\tau_{\rm line}$, this allows a direct estimate
of the (isotropic) amount of iron in the absorber. 

It came as a big surprise that these estimates, applied to the 
parameters of the transient absorption line in GRB~990705\cite{amati00}, 
yielded an estimate of $M_{\rm Fe} \sim 44 \, \Omega \, M_{\odot}$
within $R \lesssim 1.3$~pc, where $\Omega$ is the solid angle covered 
by the absorber as seen from the GRB source (see the corresponding
fits in Fig.~\ref{fig990705}). Since this does obviously not seem
realistic in any known astrophysical setting, additional effects due 
to clumping of the absorber in small clouds, in which recombination
would become efficient\cite{boettcher01}, or resonance scattering 
of the Fe~XXVI~Ly$\alpha$ line out of the line of sight\cite{lazzati01}
had been considered. Both of these effects could plausibly reduce 
the necessary amount of iron in the absorber to $M_{\rm Fe} \lesssim 
1 \, M_{\odot}$, but require a rather extreme degree of clumping, 
with densities of $n \sim 10^{11}$~cm$^{-3}$ in the clumps and 
distance/size ratios of $x/r \sim 10^3$ -- $10^5$. 

B\"ottcher et al.\cite{bfd02} have scaled the required parameters of the 
absorber to the clumping properties of supernova ejecta, derived from 
detailed 3-D hydrodynamics simulations of supernovae. The results
were parameterized in terms of the total mass contained in the dense
absorbing clouds, $M_{\rm cl}$, and the time delay $\Delta t$ between 
the supernova producing the absorbing ejecta, and the GRB. In the
case of a supranova scenario, $\Delta t$ is the delay between the
progenitor supernova and the GRB, while in the He-merger scenario,
$\Delta t$ represents the time between the primary's supernova
explosion and the He-merger-triggered GRB. Other progenitor models,
such as the collapsar/hypernova models, would possess too dilute
environments in order to be consistent with the observed properties
of GRB~990705. The results of B\"ottcher et al.\cite{bfd02} are 
summarized in Fig.~\ref{abs_parameters}, and illustrate that all 
currently discussed GRB models seem to be hard-pressed to produce 
the required environments to reproduce the transient absorption 
feature in GRB~990705, with the possible exception of the supranova 
scenario.

\section{\label{discussion}Summary and Discussion}

In this review, I have presented a comprehensive overview of the
observed X-ray emission line features in early GRB afterglows and
the singular case of a transient X-ray absorption edge during the
prompt phase of a GRB. So far, 4 cases of $\sim 3 \, \sigma$ iron
line detections have been found, and 3 cases of lower-energy X-ray
emission line features, consistent with K$\alpha$ emission from
lighter metals without clearly detectable Fe-group emission lines. 
Models to explain the observed X-ray emission line features can be 
grouped into photoionization (reflection) and thermal models; 
reflection models can be further subdivided into distant and nearby 
reprocessor scenarios. 

Generally, the observed Fe K$\alpha$ fluorescence or recombination
lines could be reasonably well represented by both photoionization 
and thermal models, though they tend to require rather extreme
parameters in terms of iron abundance and total amount of iron
in the line-emitting material and/or geometry. 

If the observations of K$\alpha$ lines of lighter elements without 
iron-group element fluorescence/recombination lines are real and
not related to abundance effects, they might strongly support thermal
emission models because any reflection model based on an incident continuum 
extending into the hard X-ray regime, would naturally also produce strong 
Fe-group element lines. The suppression of Fe-group lines in reflection 
models would require a rather unlikely degree of fine-tuning of parameters, 
e.g., a very soft incident X-ray spectrum and/or an ionization parameter 
very close to $\xi \sim 100$~erg~cm~s$^{-1}$ (see Ref.~\refcite{lazzati02}).

One peculiar observation about the presence or absence, respectively of
X-ray emission lines in early GRB afterglows seems to be in order here.
Tab.~\ref{line_table} shows that the first 4 cases of line detections (until
GRB~000214) all only detected lines consistent with Fe K$\alpha$ emission;
no other lines were found. Intriguingly, the later 3 cases (GRB~011211,
GRB~020813, and GRB~030227) all exclusively detected lines of lower-Z
elements, but did not find significant evidence for emission lines of
iron group elements. This may be in part due to the specific instrument
capabilities. {\it BeppoSAX} had very limited spectral resolution at
energies below a few keV and would probably not have been able to detect
the low-Z element emission lines found in the later bursts. They
were completely outside the sensitive energy range of {\it ASCA},
and the {\it Chandra} HETG observation of GRB~991216 was also clearly
optimized for line detections around the iron-group element K$\alpha$
complex. However, the non-detection of any Fe-group line in any GRB 
afterglow after 000214 --- in spite of many rapid follow-up observations 
by {\it XMM-Newton} and {\it Chandra} --- may raise some concern 
about the credibility of the previous detections. An interesting
observation in this regard, however, is that all the GRBs with reported
Fe-group line detections had hosts at relatively low red shifts of 
$z \le 1$. In contrast, the later detections of lighter-element K$\alpha$ 
emission lines were all associated with GRBs at $z > 1$ (assuming that
the tentative identifications of the lines in GRB~030227 are correct). 
Also, the low-Z element emission lines in the low-red-shift GRBs all
seem to be associated with bulk outflows with velocities $v \gtrsim 0.1$~c, 
while the iron-group line detections indicated much smaller outflow
velocities, up to $v \sim 0.05$~c in the case of GRB~991216\cite{piro00}.

These observations are clearly still based on very-low-number ``statistics''
and need further confirmation. However, if confirmed, they may reveal
information about the cosmological evolution of GRB progenitors. Lower
metallicity stars at higher red shifts might suffer less mass loss
during their main-sequence life time than stars of the same initial
mass during the present cosmological epoch. Consequently, the initial
mass threshold for producing a GRB might have been lower in the early
Universe than it is at lower red shifts. During a supernova explosion
probably associated with a GRB, the outflowing, shocked shells of the
progenitor's envelope would encounter a lower external density, 
resulting in a higher bulk velocity out to large distances. If 
the observed bulk velocity is indeed related to the motion of a
pre-GRB supernova explosion --- as in the supranova model ---, this
would at least qualitatively explain the observed trends. Extrapolating 
to very high red shifts, this is consistent with the hypothesis that 
high-red-shift GRBs may be associated with the almost metal-free 
(pop. III) earliest populations of stars in the Universe.

%
%
%
%

\end{document}